\documentclass[prl,preprint,showpacs,floatfix,amsmath,amssymb,superscriptaddress]{revtex4-1}
\usepackage{amsfonts}
\usepackage{stmaryrd}
\usepackage{bbm}
\usepackage{mathrsfs}
\usepackage{tipa}
\usepackage{amssymb}
\usepackage{txfonts}
\usepackage{graphicx}
\usepackage{dcolumn}
\usepackage{epstopdf}
\usepackage[colorlinks,linkcolor=blue,urlcolor=blue,citecolor=blue]{hyperref}
\usepackage{multirow}
\usepackage{subfigure}
\usepackage{url}
\usepackage{upgreek}
\usepackage[utf8]{inputenc}
\usepackage[english]{babel}

\begin{document}

\title{Revealing strong coupling of collective modes between superconductivity and pseudogap in cuprate superconductor by terahertz third harmonic generation}

\author{J. Y. Yuan}
\affiliation{International Center for Quantum Materials, School of Physics, Peking University, Beijing 100871, China}

\author{L. Y. Shi}
\affiliation{International Center for Quantum Materials, School of Physics, Peking University, Beijing 100871, China}

\author{L. Yue}
\affiliation{International Center for Quantum Materials, School of Physics, Peking University, Beijing 100871, China}

\author{B. H. Li}
\affiliation{Beijing academy of quantum information science, Beijing, 100193, China}

\author{Z. X. Wang}
\affiliation{International Center for Quantum Materials, School of Physics, Peking University, Beijing 100871, China}

\author{S. X. Xu}
\affiliation{International Center for Quantum Materials, School of Physics, Peking University, Beijing 100871, China}

\author{T. Q. Xu}
\affiliation{Applied Superconductivity Center and State Key Laboratory for Mesoscopic Physics, School of Physics, Peking University, Beijing 100871, China}

\author{Y. Wang}
\affiliation{Applied Superconductivity Center and State Key Laboratory for Mesoscopic Physics, School of Physics, Peking University, Beijing 100871, China}
\affiliation{Peking University Yangtze Delta Institute of Optoelectronics, Nantong, Jiangsu, China}

\author{Z. Z. Gan}
\affiliation{Applied Superconductivity Center and State Key Laboratory for Mesoscopic Physics, School of Physics, Peking University, Beijing 100871, China}

\affiliation{Peking University Yangtze Delta Institute of Optoelectronics, Nantong, Jiangsu, China}

\author{F. C. Chen}
\affiliation{Beijing National Laboratory for Condensed Matter Physics, Institute of Physics, Chinese Academy of Sciences, Beijing 100190, China}

\author{Z. F. Lin}
\affiliation{Beijing National Laboratory for Condensed Matter Physics, Institute of Physics, Chinese Academy of Sciences, Beijing 100190, China}

\author{X. Wang}
\affiliation{Beijing National Laboratory for Condensed Matter Physics, Institute of Physics, Chinese Academy of Sciences, Beijing 100190, China}

\author{K. Jin}
\affiliation{Beijing National Laboratory for Condensed Matter Physics, Institute of Physics, Chinese Academy of Sciences, Beijing 100190, China}

\author{X. B. Wang}
\affiliation{Beijing National Laboratory for Condensed Matter Physics, Institute of Physics, Chinese Academy of Sciences, Beijing 100190, China}

\author{J. L. Luo}
\affiliation{Beijing National Laboratory for Condensed Matter Physics, Institute of Physics, Chinese Academy of Sciences, Beijing 100190, China}

\author{S. J. Zhang}
\affiliation{International Center for Quantum Materials, School of Physics, Peking University, Beijing 100871, China}

\author{Q. Wu}
\affiliation{International Center for Quantum Materials, School of Physics, Peking University, Beijing 100871, China}

\author{Q. M. Liu}
\affiliation{International Center for Quantum Materials, School of Physics, Peking University, Beijing 100871, China}

\author{T. C. Hu}
\affiliation{International Center for Quantum Materials, School of Physics, Peking University, Beijing 100871, China}

\author{R. S. Li}
\affiliation{International Center for Quantum Materials, School of Physics, Peking University, Beijing 100871, China}

\author{X. Y. Zhou}
\affiliation{International Center for Quantum Materials, School of Physics, Peking University, Beijing 100871, China}

\author{D. Wu}
\affiliation{Beijing academy of quantum information science, Beijing, 100193, China}

\author{T. Dong}
\email{taodong@pku.edu.cn}
\affiliation{International Center for Quantum Materials, School of Physics, Peking University, Beijing 100871, China}

\author{N. L. Wang}
\email{nlwang@pku.edu.cn}
\affiliation{International Center for Quantum Materials, School of Physics, Peking University, Beijing 100871, China}
\affiliation{Beijing academy of quantum information science, Beijing, 100193, China}
\affiliation{Collaborative Innovation Center of Quantum Matter, Beijing, China}

\date{\today}

\begin{abstract}

The study of interaction between different degrees of freedom in solids is of fundamental importance to understand the functionalities of materials. One striking example of such interaction is the intertwined coupling or competition between superconductivity (SC), charge density wave (CDW), pseudogap state (PG), and other exotic phases in cuprate superconductors. Recent emergence of nonlinear Terahertz (THz) third harmonic generation (THG) spectroscopy provides a powerful tool for exploring the collective (Higgs) modes of superconductivity order parameters, and its interaction with intertwined/competing phases. In this study, we report on nonlinear THz THG spectroscopy of the YBa$_2$Cu$_3$O$_{6+x}$ (YBCO) thin films with different doping. We identify a characteristic temperature $T_{THG}$, below which third order suscepetility $\chi^{(3)}$ emerges. Notably, the $T_{THG}$ is coincident with the crossover temperature $T^*$ of pseudogap in a wide range doping of phase diagram. Upon entering the superconducting state, THG increases sharply but exhibits an abnormal dip feature near $T_c$ which is more clearly seen in optimally doped sample. Strikingly, we observe a beating structure directly in the measured real time waveform of THG signal. Fourier transformation of the time domain waveform gives two separate modes below and above original THG frequency. The observation strongly indicates that an additional mode, presumably Higgs mode, appears at $T_c$ and couples to the mode already developed below $T^*$. The strong coupling effect offers new insight into the interplay between superconductivity and pseudogap. The result unambiguously suggests that the pseudogap phase is not a precursor of superconductivity but represents a distinct order.

\end{abstract}

\maketitle

The search for the underlying mechanism of superconductivity and its interplay with pseudogap state is at the heart of cuprate superconductors \cite{cuprate_review1}. A long time controversial issue in the quest to understand the superconductivity in cuprates is whether the pseudogap state is a precursor to macroscopic coherent superconductivity or a distinct phase characterized by broken symmetries below the onset temperature $T^*$. Typical linear spectroscopic techniques or measurement under equilibrium state for a superconductor are only sensitive to the single particle excitation, enabling determination of energy gaps developed in superconducting or pseudogap phases and characterization of different physical properties \cite{Cuprate_review_ARPES}. By contrast, the collective excitations of superconductors can not be accessed by such measurements, except for some special cases \cite{PhysRevLett.45.660,PhysRevB.89.060503}. The superconductivity breaks $U(1)$ gauge symmetry and the associated collective modes are well understood. While the Goldstone mode is shifted to plasma energy by Anderson-Higgs mechanism \cite{anderson1,Anderson2}, the Higgs mode locates at the energy corresponding to the minimum of the single particle excitation spectrum with energy of $2\Delta$ \cite{Littlewood1981,Varma2002,Higgs2}. The generation of sufficiently strong THz pulse \cite{LNO_THz_generation} have enabled the direct access to Higgs mode in superconductors via nonlinear optical coupling effect in the form of either free or forced oscillations in time domain \cite{shimano_NbN_PRL,Matsunaga_NbN_Science,Higgs2,Higgs_shimano_NbN_polarization,JGWang_NPhoto19}. Successful observation of superconductivity contributed THz third harmonic generation (THG) signal in conventional superconductors has raised the hope for study of the interaction of different orders in dynamical fashion via coherent THz nonlinearities. Cuprate superconductor is an ideal playground for exploring the dynamical interaction between Higgs mode and other distinct collective modes \cite{cuprate_higgs_theory1,cuprate_higgs_theory2,cuprate_higgs_theory3}. As cuprate systems have very complex phase diagrams, the interaction between superconductivity and other intertwined orders may modify the spectroscopic feature of the Higgs oscillation. The reported studies of Higgs mode by THz nonlinear Kerr \cite{Higgs_Bi2212_PRL,Higgs_Bi2212_doping_dependent} and THz THG spectroscopy \cite{Chu20} in cuprates have generally concentrated on determining which channel (Higgs mode \cite{ShimanoTsuji20} or Cooper pair breaking \cite{lara_CDF1}) dominates the nonlinear THG and its temperature evolution. The interaction between Higgs mode and other distinct collective mode induced anti-resonance in temperature dependence of THG intensity has been reported \cite{Chu20,Fano_Higgs_2022}. However, the intertwined order coupling effect on Higgs oscillation waveform in cuprate superconductors remains largely unexplored.

In this manuscript, we report a systematic nonlinear THz THG study on four different doping YBCO thin films driven by multicycle strong filed THz pulse (see methods for additional experimental details). We identify a characteristic temperature $T_{THG}$ below which the third order susceptibility $\chi^{(3)}$ starts to grow, and find that $T_{THG}$ matches well with pseudogap developing temperature $T^*$. Upon entering the superconducting state, $\chi^{(3)}$ increases sharply but exhibits an abnormal dip feature near $T_c$ which is more clearly seen in optimally doped sample. Surprisingly, we observe a beating structure directly in the measured real time waveform of THG signal. Fourier transformation of the time domain spectrum results in two separate modes below and above original THG frequency. The observation strongly indicates that an additional mode, presumably Higgs mode, appears at $T_c$ and couples to the mode already developed below $T^*$. We also perform similar measurement on an electron-doped cuprate sample La$_{2-x}$Ce$_x$CuO$_4$ (x$\sim$0.15) (LCCO) without pseudogap phase and confirm the absence of THG above its $T_c$. The striking coupling effect offers new insight into the interplay between superconductivity and pseudogap.

\begin{figure}
\includegraphics[width=15cm]{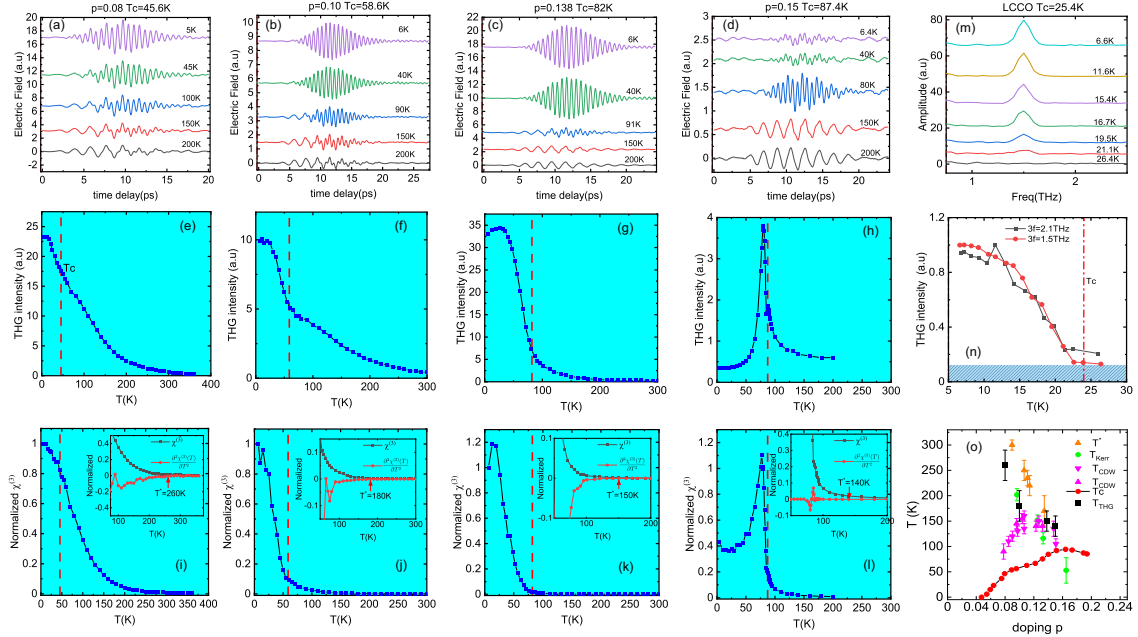}
\caption{\label{Fig1}
(a)-(d), The real time waveforms of the recorded THG signal after a high frequency(3$\omega$=1.5THz) bandpass filter at selected temperatures for (a) p=0.08, (b) p=0.10, (c) p=0.13, (d) p=0.15 YBCO thin films; (e)-(h), Temperature dependence of $I_{3\omega}$ from all four samples; The red dash lines label the $T_c$ of the samples. (i)-(l), Temperature dependence of normalized third order susceptibility $\chi^3(T)$ from all four samples, the insets show the second derivative of $\chi^3(T)$ with respect to temperature $\frac{\partial^2 \chi^3(T)}{\partial T^2}$. The red vertical arrows denote the crossing temperature of the rapid growing of the THG signals $T_{THG}$ determined from the second derivative. (m) The THG spectrum of optimal doped LCCO with Tc=24K driven by 0.5 THz pulse. (n) Temperature dependence of $I_{3\omega}$ for LCCO sample. The THG intensity disappears abruptly when the temperature is higher than $T_c$. (o) Temperature-doping phase diagram of YBCO. Black squares represent the crossing temperate $T_{THG}$ determined by the present THz THG spectroscopy. For comparison, the pseudogap temperatures determined by other probes are also plotted. Orange triangles are $T^*$ reported by polarized neutron scattering\cite{neutron_T_PG}. Magenta triangles represent the formation temperature of the short-range CDW, $T_{CDW}$, reported by resonant X-ray measurements\cite{YBCO_CO_RIXS_2012,YBCO_CDW_dome_2014}. Green dots are the temperature $T_{Kerr}$ below which the time reversal symmetry is broken, reported by the polar Kerr effect\cite{Kerr_Xia_2008}. Red dots are the superconductivity transition temperate $T_c$, reported by resistivity measurements\cite{YBCO_SC_Liang_2006}.
}
\end{figure}

We start with the temperature dependence of THG intensity for four different doping level samples driven by 60 kV/cm 0.5 THz multi-cycle pulse. The waveform and electric field of the fundamental pulse are shown in supplementary materials. Fig.~\ref{Fig1} (a)-(d) display the typical real-time waveforms of THG signal. After fast Fourier transformation (FFT), the residual fundamental pulse after the high-frequency bandpass filter (1.5 THz) and the THG signal can be well separated in the frequency domain. To evaluate the temperature dependence of the THG signal, the spectral weight from 1.25 THz to 1.75 THz is integrated. As shown in Fig.~\ref{Fig1} (e)-(h), the integrated THG intensity decreases continuously with an increasing temperature towards $T_c$ expected for the optimal doping, which represents a resonant enhancement of THG signal at selective temperature. We will discuss the nature of resonance later. Above $T_c$, the THG signal still survive up to high temperature, which is in contrast to the conventional superconductors \cite{Matsunaga_NbN_Science, MgB2_Leggett_TPTR,Higgs_NbN_ZXWang,Higgs_MgB2_DT} and iron-based superconductors \cite{Higgs_Fe_single,Higgs_FeTeSe}. For p=0.15, 0.13, and 0.10 doping levels, a substantial enhancement of the THG signal below $T_c$ is observed, whereas, for deeply underdoped p=0.08, only a small upturn of the THG intensity is seen. The temperature dependence of raw THG intensity should be affected by the effective electric field inside the thin film because the effective driving field inside the thin film $E_{\omega}(T)$ decreases below $T_c$ due to the increase of the reflectivity. To quantitatively evaluate the temperature dependent behavior of the THG signal and get an innate response of THG efficiency, we calculated the third-order nonlinear susceptibility $\chi^{(3)}(T)$ for all of the doping samples driving by 0.5 THz. The $\chi^{(3)}(T)$ is calculated by simultaneous measurement of the THG and the residual fundamental frequency passing through one band-pass filter of $3\omega$: $\chi^{(3)}=E_{3\omega}/E_{\omega}^3$. The comparison between the above approximation and the calculation from the Fabry-Perot formula further illustrates the rationality of above analysis, as shown in the supplementary materials. Fig.~\ref{Fig1} (i)-(l) show the normalized $\chi^{(3)}$ as a function of temperatures for p=0.08, 0.10, 0.13, and 0.15, respectively. The resonant behavior is still visible for optimal doping after taking account of the screening effect, which is distinct from the reported measurements \cite{Chu20}. We will discuss it later. For p=0.10, 0.13 thin films, $\chi^{(3)}$ increase monotonically as the temperature decrease from 300 K and shows a clear enhancement around $T_c$. However, $\chi^{(3)}$ does not show anomaly at $T_c$ for p=0.08 sample. To quantitatively define a cross temperature of the emerging THG signal, we follow the method using in ref. \cite{Higgs_Bi2212_PRL} to take the second derivative of $\chi^{(3)}$ concerning temperature as shown in the insets of Fig.~\ref{Fig1} (i)-(l). The onset temperature $T^{*}_{THG}$ is determined as the slope change in the temperature dependence of the second derivative. We extract the $T^{*}_{THG}$ = 260 K for p=0.08, 180 K for p=0.10, 150 K for p=0.13 and 140 K for p=0.15.

We summarize the extracted crossover temperature $T^{*}_{THG}$ in Fig.~\ref{Fig1}(o), in which the reported characteristic temperatures for the exotic phase are also displayed. Obviously, the $T^{*}_{THG}$ follows the tendency of pseudogap temperature line. In order to provide a strong experimental basis, we perform additional experimental measurements in support of this correlation. First, it is well accepted that the pseudogap phase is present in the hole-doped side but absent in the electron-doped side in the phase diagram of cuprate. For a comparison, we have done similar measurement on an electron-doped cuprate sample La$_{2-x}$Ce$_x$CuO$_4$ (x$\sim$0.15) (LCCO) with $T_c$=25 K and confirm the absence of THG above its $T_c$. The THG signals at selective temperatures on LCCO driven by f=0.5 THz and f=0.7 multicycle THz pulses are shown in Fig.~\ref{Fig1}(m) and (n). The THG signal disappears abruptly when the temperature is heated up to $T_c$, similar to the THG response of conventional superconductors\cite{Matsunaga_NbN_Science, MgB2_Leggett_TPTR, Higgs_MgB2_DT, Higgs_NbN_ZXWang} and iron-pnictides superconductors\cite{Higgs_Fe_single, Higgs_FeTeSe}. This comparison highly suggests that the THG signal above $T_c$ is linked to the exotic pseudogap state in hole doped cuprate superconductors. Secondly, although the third order nonlinear process is symmetry allowed in all of the media as long as the driving pulse is strong enough, the observed behavior is different from the contribution of normal quasiparticles without displaying exotic properties. To this purpose, we tried to detect the THG signal driven by strong near-infrared (NIR) pulse with the stimulating electric field about ten times stronger than the THz driving pulse. As shown in supplementary file, although we can resolve the THG signal with NIR driving, the THG intensity in (S, S) geometry exhibits a characteristic fourfold symmetry expected from the $D_{4h}$ lattice symmetry, likely originated from the intraband acceleration of quasiparticles or interband polarization \cite{HHG_solid_review}. In contrast, the THz THG intensity is essentially isotropic as we shall present below.

With these observations, we suggests that it is the pseudogap state that gives rise to a considerable normal state THG signal in hole doped cuprate superconductors. However, what kind of order in the pseudogap phase that leads to the THG process remains a mystery. The issue is linked to the origin of the pseudogap for which there have been many different suggestions such as preformed Cooper pairs, 2D or fluctuated CDW order, etc. The implication of our experiments on the origin of the pseudogap will be discussed later.

\begin{figure}
\includegraphics[width=16cm]{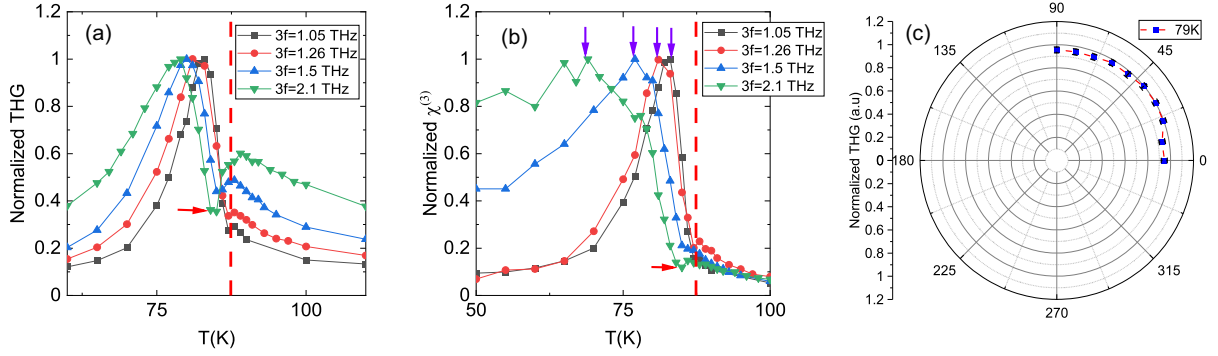}
\caption{\label{Fig2}
(a) Temperature dependence of I$_{3\omega}$  in optimal doped YBCO thin film driven by $\omega$=0.35, 0.42, 0.5, 0.7 THz multicycle pulse; The dash line indicates the superconductivity transition temperature $T_c$=87.4K; The red arrow labels the THG intensity dip at different temperate varying with the frequency of the driven pulse; (b) Temperature dependence of $\chi^{(3)}$ from all driven THz pulses plotted with normalized scale. The resonance is still visible after considering the screening effects, suggesting the Higgs mode is dominating the THG signal. The dip feature still survives after taking account of the screening effect, as labeling by a red arrow. (c) The polarization dependence of I$_{3\omega}$ with (S, S) geometry driven by 0.5 THz pulse. The sample is rotated with respect to the polarization co-plane of fundamental and 3$\omega$ photons by a low-temperature piezo rotator. The isotropic response indicates the Higgs mode dominates the THG signal.
}
\end{figure}

Having established the connection of the crossing temperature of THG with the pseudogap state, we proceed to study the interaction between the pseudogap and superconducting order parameters. To this end, we examine THG measurement on the optimal doping sample (OP87) with $T_c=87 K$, of which the superconducting transition is sharp, suggesting the homogeneity of the superconductivity phase. We first discuss the origin of the superconducting state THG signal, which is essential to study the interaction. Systematic measurements with selective central frequencies were done to further confirm the universal resonance presented in 0.5 THz driving. As shown in Fig.~\ref{Fig2} (a), the THG intensity resonance appears at all the driving frequencies, of which resonance temperature decreases with increasing driving frequency. After taking into account the screening effect, as shown in Fig.~\ref{Fig2} (b), the resonance still survives at all driving frequencies. The presence of resonance when twice the pump-pulse frequency matches the superconducting gap frequency and the isotropic THG intensity with respect to the lattice axis are two key ingredients for identifying the Higgs mode dominating THG signal in experiment \cite{Matsunaga_NbN_Science,Higgs_shimano_NbN_polarization}. In cuprates, it is difficult to match the resonance temperature with the gap function due to the presence of nodal in the gap function in which the gap amplitude varies from zero at the the nodal direction to maximum value along the anti-nodal direction continuously. Nevertheless, the survived resonance suggests that the Higgs mode predominates the THG signal at low temperatures. Further support for the assignments comes from the polarization dependence measurement. As mentioned, the THG response of Higgs mode and Cooper pair breaking or charge density fluctuation process have different polarization dependence. The first one is isotropic, while the latter is anisotropic, following the underlying lattice symmetry \cite{Higgs_shimano_NbN_polarization}. Fig.~\ref{Fig2} (c) shows the polarization scan with respect to the lattice axis with (S,S) geometry at the corresponding temperature for resonance. The THG intensity is isotropic within the experimental error. With these observations, we suggest THG signal in the superconducting state is dominated by the Higgs mode of superconductivity order parameter rather than the Cooper pair breaking. It is worth noting that the resonance behavior disappears for the p=0.13 doping sample with $T_c$=82 K. On the other hand, the reported THG measurement on different families of optimal hole-doped cuprates does not show resonance after considering the screening effect \cite{Chu20}. These results may indicate that disorder plays an important role in the detection of Higgs mode in cuprate superconductors \cite{cuprate_THG_disorder_Lara,cuprate_THG_disorder_Lara2}.

\begin{figure}
\includegraphics[width=16cm]{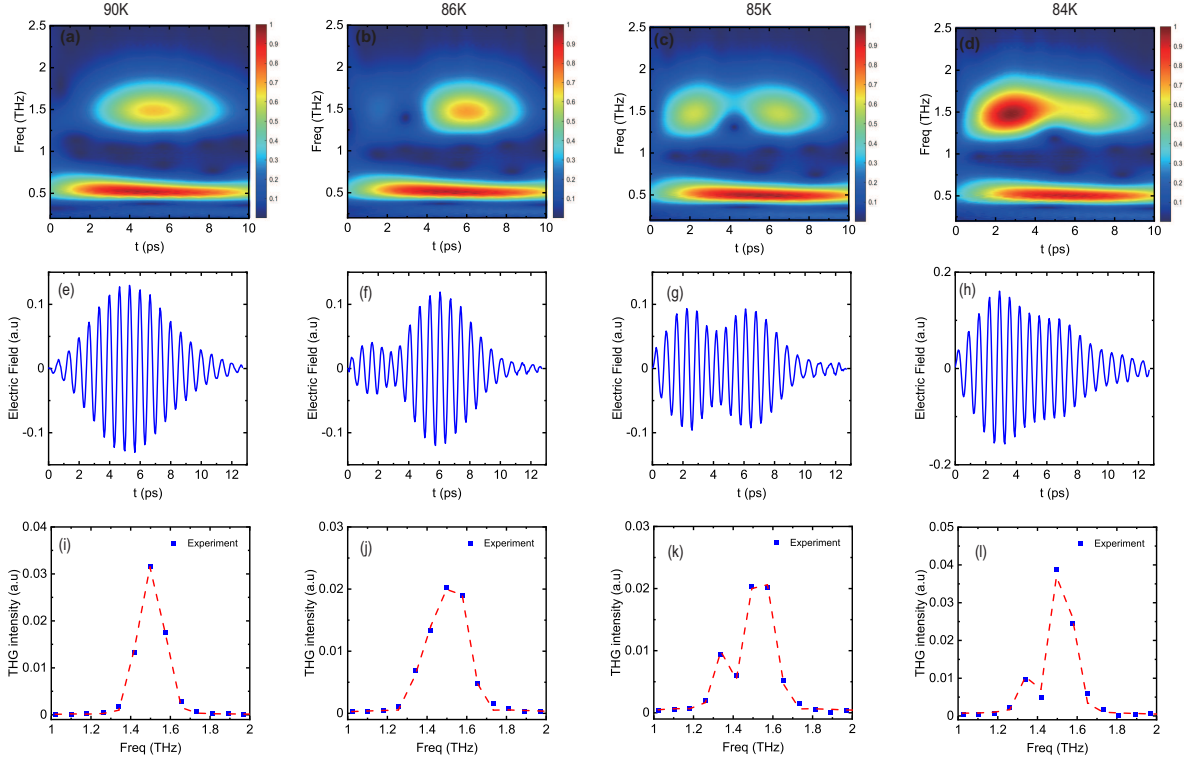}
\caption{\label{Fig3}
(a)-(d), THG intensity continuous wavelet transformation (CWT) chronograms in optimal doped YBCO thin film driven by 0.5 THz pulse for (a) 89 K, (b) 86 K,(c) 85 K and (d) 84 K. The dip at selective delay time in CWT chronograms indicates a destructive interference. (e)-(h) The real time waveforms of THG signal after a digital 1THz high pass filter. The beating pattern in the waveform is corresponding to the dip in CWT chronogram. (i)-(l) The THG spectrum for 89 K, 86 K, 85 K and 84 K after the global FFT. The splitting is caused by the dynamical strong coupling between Higgs mode and pseudogap collective mode as described in main text.
}
\end{figure}

We turn to study the superconductivity-pseudogap interplay by recalling the THG intensity minimum at 85 K by 0.5 THz driving.  As shown in Fig.~\ref{Fig2} (a), the THG intensity dip appears at all the driving frequencies, whose dip temperature decreases with increasing driving frequency. Notably, the dip feature still survives, although the sharp dip is smeared out after taking account of the screening effect, as shown in Fig.~\ref{Fig2} (b). This observation suggests that the THG in YBCO below Tc can not be described by assuming that only superconducting condensate contributes to the THG signal. Otherwise, as observed in conventional superconductors, the THG intensity should grow monotonically with decreasing temperature before a resonance appears. To reveal the underlying origin of the dip at the corresponding temperate, we examine the waveforms of the THG signal in time domain and corresponding spectra in frequency domain more closely nearby $T_c$ for four different temperatures 90, 86, 85, 84 K, as shown in Fig.~\ref{Fig3}. Fig.~\ref{Fig3} (a)-(d) show the time-dependent spectral response obtained by continuous wavelet transformation (CWT). At temperature away from $T_c$, e.g. 90 K above $T_c$, the THG signal is contributed only from the pseudogap phase, we find a single elliptical structure along the delay time axis for both fundamental and THG pulses. However, the time-frequency THG intensity mappings in Fig.~\ref{Fig3} (b), (c), and(d) are found to have the obvious dip at 3 ps for 86 K, 4.5 ps for 85 K, and 5.5 ps for 84 K, respectively. Presence of the dip structure coincides with the dip in the integrated THG intensity as a function of temperature displayed in Fig.~\ref{Fig2} (a). Fig.~\ref{Fig3} (e)-(h) show the real-time THG waveforms after a digital 1 THz high pass filter for the four different temperatures. Strikingly, a beating pattern emerges on the THG oscillations, which is corresponding to the dip in the time-frequency THG intensity mapping. The development of beating indicates the splitting of the THG frequency, which is indeed seen clearly in the fast Fourier transformation (FFT) spectra, as shown in Fig.~\ref{Fig3} (i)-(l). As can be resolved from the plot, one of the split two peaks is unambiguously pushed down to a lower frequency below 1.5 THz, and the other to a higher frequency above 1.5 THz.

Now we attempt to identify the nature of the observed beating. The possibility of the Fabry-Perot effect of either the fundamental pulse or THG pulse inside the thin film introducing the experimental feature is ruled out. We record the temperature dependent waveform of a 1.5 THz multicycle pulse in linear response regime passing through the thin film. A continued phase shift induced by superconductivity is observed, but no beating pattern is identified around $T_c$. Then, the beating would mean that the strongly coupled modes meet at the same energy and result in a level repulsion. It is important to note that a collective mode is present in the the pseudogap phase, which results in the THG with strong multicycle THz pulse driving. Upon entering the superconducting state, an additional mode appears at $T_c$ and couples to the mode already developed in the pseudogap phase. We have elaborated above that this new mode developed below $T_c$ should be collective Higgs mode of superconductor. Since both modes are driven by the same strong field THz fundamental pulse, they should have the same energy. However, because they have different origins, their coupling leads to the energy splitting.

\begin{figure}
\includegraphics[width=14cm]{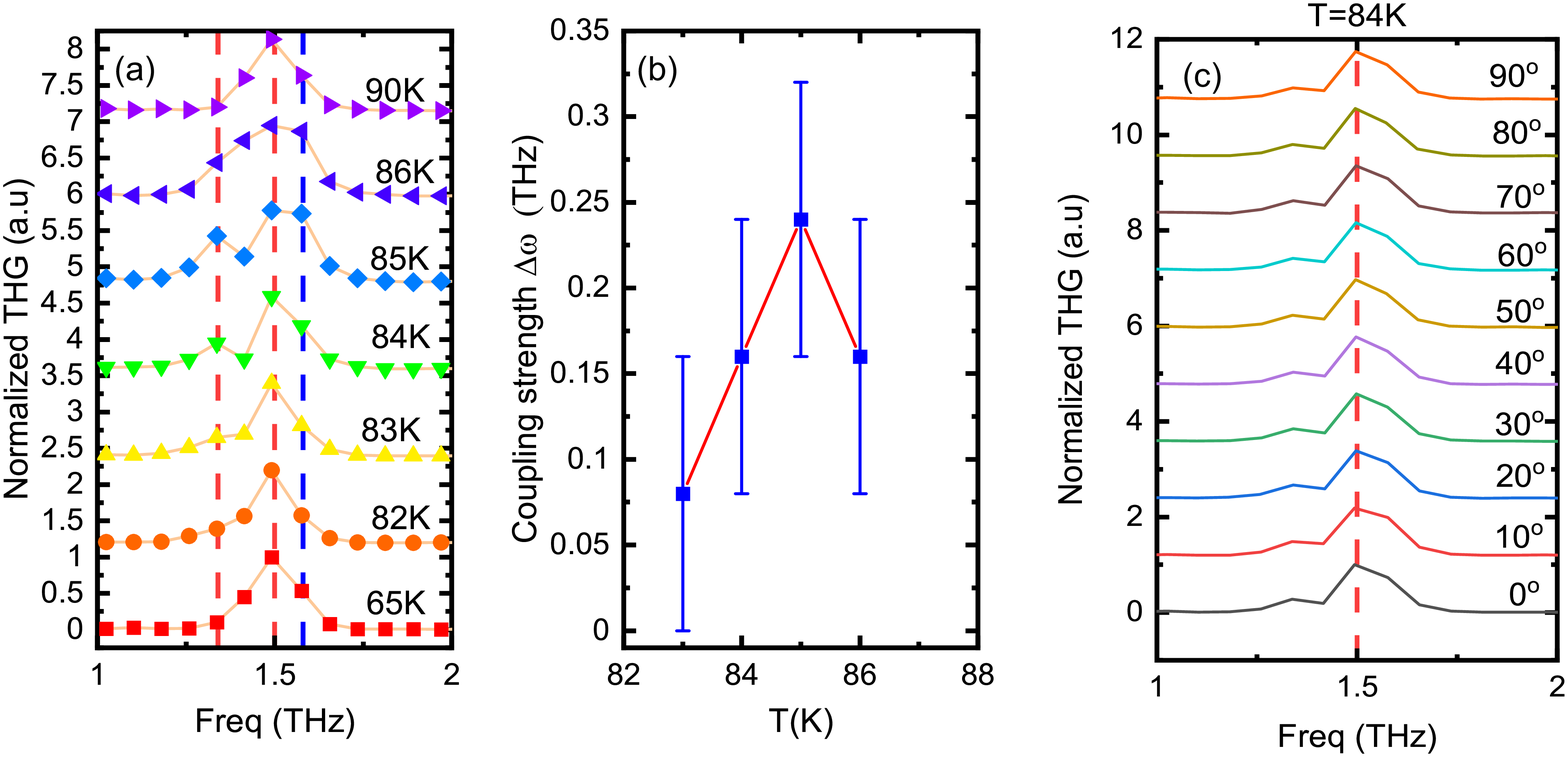}
\caption{\label{Fig4}
 (a) Temperature dependence of the profile of THG spectra. The lineshape is change dramatically when the temperature is decreased, first as a Lorentz profile above $T_c$, then as a double peak around 85 K, subsequently an asymmetric peak in the intermediated temperature (82 K), and finally as a symmetric Lorentz peak again below 65 K. The red and blue dash line indicate the new peak at lower frequency side, and spectral enhancement at high frequency part, respectively. (b) Temperature dependence of extracted splitting width from (a); The peak of splitting at 85 K indicates a dynamical enhancement of coupling between Higgs mode and pseudaogap collective mode. (c) The isotropic response of the hybridization, which is consistent with the isotropic THG response of the Higgs mode and pseudogap phase.
}
\end{figure}

To resolve the evolution of peak splitting more clearly, we analyze the spectral feature in more detail. Fig.~\ref{Fig4} (a) shows the evolution of normalized THG spectra as a function of temperature. The lineshape is seen to change dramatically when the temperature is decreased, first as a Lorentz profile above $T_c$, then as a double peak around 85 K, subsequently an asymmetric peak in the intermediated temperature (82 K), and finally as a symmetric Lorentz peak again below 65 K. The spectral shape evolution reflects the coupling strength and possible spectral weight transfer between different modes. Fig.~\ref{Fig4} (b) displays the extracted splitting width of THG spectra in a narrow temperature window $\Delta\omega$ which characterizes the coupling strength. The $\Delta\omega$ is peaked at 85 K with $\Delta\omega$ =0.25$\pm$ 0.08 THz, indicating the dynamical enhancement of coupling strength when the resonant driving of the Higgs mode of the superconductivity order parameter is nearly fulfilled. The enhanced coupling dynamically brings the coupled system into the very strong coupling regime. The resulted strong interaction then repels the two bare oscillations towards lower and higher frequency, respectively. The ratio of spectra splitting and the bare frequencies $\Delta\omega/\omega$ reaches 16 $\%$, which is a remarkably large value. Fig.~\ref{Fig4} (c) displays the polarization dependence of the hybridization, which is isotropic with respect to the lattice axis. Such behavior is not unexpected since both the THG signals arising from Higgs mode and pesudogap phase appear to be isotropic.

Our new experimental results have profound implication for further understanding the interplay between superconductivity and pseudogap. As mentioned before, the pseudogap phase in hole-doped cuprates is characterized as a depletion of the single-particle spectral weight near the chemical potential at a crossing temperature $T^*$. At its heart is whether this pseudogap arises from precursor superconductivity or a distinct order parameter. The beating behavior of the THG signal nearby $T_c$ between superconductivity and pseudogap indicates that the pseudogap phase is a distinct state rather than a precursor of superconductivity. The reason is rather obvious: if the pseudogap phase is uncoherent Cooper pairs, once the temperature lowers down to $T_c$, the macroscopic coherence of Cooper pairs will establish a single superconductivity phase. Then there leaves one path for generating the THG signal, and the beating pattern in real time waveform of THG signal will not appear. Although for the underdoped samples, we do not observe a clear beating pattern nearby $T_c$, it is likely to attribute to the increasing inhomogeneity of sample quality. Our results reveal a dynamical strong coupling between Higgs mode and pseudogap collective mode in a driving protocol. However, the intrinsic damping rate and the center frequency of the two coupled modes are missing. The single cycle THz pump-THz probe in the two dimension scanning geometry is desirable to address the above question.

In the normal state, our systematic doping dependence measurement establishes the connection of onset temperature of the THG signal with the reported $T^*$. Yet, the microscopic mechanism of how the electrons in the pseudogap phase generate relatively high-efficiency THG is still missing. We notice that our extracted onset temperature $T_{THG}$ does not follow the $T_{CDW}$ dome determined by resonant inelastic X-ray spectroscopy, suggesting the normal state THG signal is not contributed by 2D-CDW order in YBCO. Beside the 2D CDW phase, there is growing evidence of a dynamical charge density fluctuation developing above the pseudogap temperature $T^*$ in a broad doping region in the hole-doped materials \cite{YBCO_CDF_2019,HBCO_CDF_2020,HBCO_CDF2_2020,CO_cuprate_review_2021}. A recent work attributed the observed THG signal above $T_c$ to the dynamical charge density fluctuation \cite{Fano_Higgs_2022}. Nonetheless, whether the normal state THG signal is solely contributed by dynamical charge density fluctuation needs further experimental exploration as well as the theoretical input. Our work open new perspective towards the exploration and manipulation of the Higgs mode-other distinct collective mode coupling via strong field nonlinear THz spectroscopy.

\begin{acknowledgments}

T. Dong gratefully acknowledges H. Chu, Y. Wan and Q. Zhang for illuminating discussions. This work was supported by the National Science Foundation of China (No.11888101, 11974414, 12134018), and the National Key Research and Development Program of China (2021YFA1400200, 2022YFA1403900, 2022YFA1603903). The linear time domain THz measurement for optimal doping YBCO (OP87) thin film was carried out at the Synergetic Extreme Condition User Facility (SECUF).

\end{acknowledgments}

\bibliographystyle{apsrev4-1}
  \bibliography{YBCO_OP_THG}

\end{document}